\documentclass[12pt,reqno]{amsart}

\usepackage{fullpage,amsmath,amssymb,amsfonts,mathrsfs,bm,graphicx,hyperref}

\def\bbone{{\mathchoice {\rm 1\mskip-4mu l} {\rm 1\mskip-4mu l}
{\rm 1\mskip-4.5mu l} {\rm 1\mskip-5mu l}}}

\newcommand{\mO}{\mathscr{O}}

\newtheorem{theorem}{Theorem}[section]

\newtheorem{lemma}{Lemma}[section]

\newtheorem{conjecture}{Conjecture}[section]

\begin{document}

\author{Abdelmalek Abdesselam}
\address{Abdelmalek Abdesselam, Department of Mathematics,
P. O. Box 400137,
University of Virginia,
Charlottesville, VA 22904-4137, USA}
\email{malek@virginia.edu}

\title{Non-Abelian correlation inequalities and stable determinantal polynomials}

\begin{abstract}
We consider the correlations of invariant observables for the $O(N)$ and $\mathbb{C}\mathbb{P}^{N-1}$ models at zero coupling, namely, with respect to the natural group-invariant measure. In the limit where one takes a large power of the integrand, we show that these correlations become inverse powers of the Kirchhoff polynomial. The latter therefore provide a simplified toy model for the investigation of inequalities between products of correlations. Properties such as ferromagnetic behavior for spin model correlations correspond, in this asymptotic limit, to log-ultramodularity which is a consequence of the Rayleigh property of the Kirchhoff polynomial.  
In addition to the above rigorous asymptotics, the main result of this article is
a general theorem which shows that inverse half-integer powers of certain determinantal stable polynomials, such as the Kirchhoff polynomials, satisfy generalizations of the GKS 2 inequalities and the Ginibre inequalities. We conclude with some open problems, e.g., the question of whether the last statement holds for powers which are not half-integers. This leads to a Hirota-bilinear analogue of the complete monotonicity property recently investigated by Scott and Sokal.
\end{abstract}

\maketitle

\tableofcontents

\section{A general setup for ferromagnetic correlation inequalities}
Let $X$ be a compact metric space and let $\mu$ be a Borel probability measure on 
$X$. We suppose we are given a collection $\mO=(\mO_1,\ldots,\mO_n)$ of continuous real-valued functions on $X$ which we call {\it basic observables}.
Throughout, our notational conventions will be such that $\mathbb{N}:=\{0,1,2,\ldots\}$ and $\mathbb{N}_{>0}:=\{1,2,3,\ldots\}$. For a multiindex $a=(a_1,\ldots,a_n)\in\mathbb{N}^n$, which we also think of as a row vector,
we will write $\mO(x)^a$ or just $\mO^a$ for the monomial $\mO_1(x)^{a_1}\cdots\mO_n(x)^{a_n}$ in the basic observables. We will use the standard terminology and notation for the length $|a|:=a_1+\cdots+a_n$ of the multiindex $a$.
For bounded Borel measurable functions $F$ on $X$ we will write, as is customary in statistical mechanics,
\[
\langle F\rangle:=\int_X F(x)\ {\rm d}\mu(x)
\]
for the expectation with respect to the measure $\mu$.

Suppose $V\in \mathbb{N}^{m\times n}$, namely, $V$ is
a $m\times n$ matrix of nonnegative integers. A collection of {\it ferromagnetic couplings} 
is a vector $J=(J_i)_{1\le i\le m}\in [0,\infty)^{m}$ which will allow us to define the unnormalized correlations
\[
\langle F\rangle_{J,{\rm un}}:=\int_X F(x)
\exp\left(\sum_{i=1}^{m}J_i \mO(x)^{V_{i\ast}}\right)
\ {\rm d}\mu(x)
\]
where $V_{i\ast}:=(V_{i,1},V_{i,2},\ldots,V_{i,n})$ denotes the $i$-th row of $V$.
The partition function is the normalization constant
$Z:=\langle 1\rangle_{J,{\rm un}}$. Since the integrand given by an exponential is continuous over a compact set, it is bounded below by its minimum and thus by a strictly positive constant. Hence $Z>0$, and
we can then define the interacting probability measure $\mu_J$ by
\[
{\rm d}\mu_J(x):=\frac{1}{Z}\exp\left(\sum_{i=1}^{m}J_i \mO(x)^{V_{i\ast}}\right)
\ {\rm d}\mu(x)\ .
\]
Expectations with respect to this 
measure will be denoted by
\[
\langle F\rangle_{J}:=\int_X F(x)\ {\rm d}\mu_J(x)=\frac{\langle F\rangle_{J,{\rm un}}}{\langle 1\rangle_{J,{\rm un}}}\ .
\]
Note that expectations $\langle\cdots\rangle$ with no $J$ subscript mean the ones with respect to the free measure $\mu$, i.e., the $J=0$ or zero coupling case of the more general family of $J$-dependent measures $\mu_J$.

\medskip\noindent
{\bf The GKS 1 inequalities:}
We will say that the system $(X,\mu,\mO)$ satisfies the GKS 1 collection of inequalities iff
$\forall m\ge 0$, $\forall V\in\mathbb{N}^{m\times n}$, $\forall J\in[0,\infty)^{m}$,
$\forall a\in\mathbb{N}^n$, we have
\[
\langle\mO^a\rangle_J\ge 0\ .
\]

\medskip\noindent
{\bf The GKS 2 inequalities:}
We will say that the system $(X,\mu,\mO)$ satisfies the GKS 2 collection of inequalities iff
$\forall m\ge 0$, $\forall V\in\mathbb{N}^{m\times n}$, $\forall J\in[0,\infty)^{m}$,
$\forall a,b\in\mathbb{N}^n$, we have
\[
\langle\mO^{a+b}\rangle_J\ge \langle\mO^{a}\rangle_J\ \langle\mO^{b}\rangle_J\ .
\]
The above setting is intentionally abstract and general, somewhat in the spirit of the presentations given in~\cite{Ginibre2,BakryE,BakryH}. For the sake of concreteness, it is helpful to keep in mind the main two examples addressed in this article, the $O(N)$ and $\mathbb{C}\mathbb{P}^{N-1}$
models, to be defined shortly. Note however, that the above formalism is much more general and applies to a wider class of models studied by physicists. The exploration of this wider class of models will be left to the forthcoming work~\cite{AbdesselamUW1,AbdesselamUW2}.

The $O(N)$ model example corresponds to taking $\Lambda$ a finite subset of $\mathbb{Z}^d$, say of cardinality $p$, and letting $X=(\mathbb{S}^{N-1})^{\Lambda}$ where $\mathbb{S}^{N-1}$ is the unit sphere in $\mathbb{R}^N$, $N\in\mathbb{N}_{>0}$.
The free measure $\mu$ is the product of copies of the unique $O(N)$-invariant Borel probability
measure on the sphere $\mathbb{S}^{N-1}$. Elements of $X$ are spin configurations $(\sigma_1,\ldots,\sigma_p)$ where we have chosen some ordering of the $p$ elements of $\Lambda$, for the sake of notational simplicity. Each spin $\sigma_i$ is a column vector $(\sigma_{i,1},\ldots,\sigma_{i,N})^{\rm T}$ in $\mathbb{S}^{N-1}$, i.e., which satisfies $\sum_{\ell=1}^N \sigma_{i,\ell}^2=1$. The action of an element $R$ of the orthogonal group $O(N)$ on a vector $\sigma\in\mathbb{S}^{N-1}$ is by matrix multiplication $(R,\sigma)\mapsto R\sigma$.
The basic observables are the inner products $\sigma_i\cdot\sigma_{i'}:=\sum_{\ell=1}^{N}\sigma_{i,\ell}\sigma_{i',\ell}$, with $1\le i<i'\le p$. Namely, $n=\binom{p}{2}$, and $\mO_1,\ldots,\mO_n$ are just some choice of ordering of the functions on $X$ given by the inner products of pairs of spins $\sigma_i\cdot\sigma_{i'}$. These are invariant observables, because they are invariant under the diagonal action of $O(N)$, namely changing all the spins by the same orthogonal matrix $R$. 
In the standard $O(N)$ model, the monomials appearing in the interaction
$\sum_{i=1}^{m}J_i \mO(x)^{V_{i\ast}}$ are single inner products, namely the $V_{i\ast}$'s are standard basis vectors of the form $e_j:=(0.\ldots,0,1,0,\ldots,0)$ with a 1 at the $j$-th entry, $1\le j\le n$.
The setting of this article allows
the inclusion of higher degree monomials, in which case we would be dealing with a generalized $O(N)$ model. Particular cases would be the $\mathbb{R}\mathbb{P}^{N-1}$ model (only $V_{i\ast}$ of the form $2e_j$) and the mixed isotensor/isovector model (a mix of $e_j$'s and $2e_j$'s), see~\cite{SokalS}.

The $\mathbb{C}\mathbb{P}^{N-1}$ model is similar, but now the spins are complex unit column vectors $z_1,\ldots,z_p$. Hence $X=(\mathbb{S}^{2N-1})^{\Lambda}$, where $\mathbb{S}^{2N-1}$ is the unit sphere of $\mathbb{C}^N$. The basic observables $\mO_j$ are given by the $\binom{p}{2}$ functions $|\langle z_i,z_{i'}\rangle|^2$, where $\langle z_i,z_{i'}\rangle:=z_i^{\ast}z_{i'}=\sum_{\ell=1}^{N}\overline{z_{i,\ell}}z_{i',\ell}$. Note that all the observables are invariant if one independently multiplies individual spins by complex numbers of modulus 1, and therefore this description is that of a lift of a probability measure which lives on a `true' $X$ given by $(\mathbb{C}\mathbb{P}^{N-1})^{\Lambda}$. The present setup also corresponds to the choice of the so-called quartic lattice action (see~\cite{RindlisbacherF} and references therein). Other options, formally equivalent in the continuum limit, are possible but we will not consider them here.
A discussion of a variety of continuum and lattice models of interest in high energy physics, including the $O(N)$ and $\mathbb{C}\mathbb{P}^{N-1}$ models as particular examples, can be found in the book by Polyakov~\cite{Polyakov}.

For many models of statistical mechanics, the GKS 1 inequalities are usually easy to prove, whereas the GKS 2 inequalities are highly nontrivial. The importance of the GKS 2 inequalities comes from the monotonicity properties one can derive as a consequence (see~\cite{PeledS} for an introductory review). For example, we have $\frac{\partial}{\partial J_i}\langle \mO^a\rangle_J=\langle \mO^{a+V_{i\ast}}\rangle_J-\langle \mO^a\rangle_J\langle \mO^{V_{i\ast}}\rangle_J$ and therefore correlations increase when the couplings $J$ are progressively turned on, as when taking an exhausting sequence of finite volumes $\Lambda$ which approximate the full lattice $\mathbb{Z}^d$. When the GKS 2 inequalities are available, this results in a robust and satisfactory way of rigorously performing the infinite volume or thermodynamic limit. After taking that limit, one can also show the correlations increase with the inverse temperature $\beta$ and the magnetic field $h$, which are factors multiplying the $J$ couplings. Note that introducing additional non-invariant observables such as $e\cdot\sigma_i$ where $e$ is a fixed preferred unit vector is easy to do within the framework of invariant observables using Griffiths' ghost spin trick~\cite{Griffiths2}. Therefore, restricting the discussion in the present article to invariant observables only is not a drastic limitation.

In the case of the $O(N)$ model, only considering invariant observables as above, the GKS 2 inequalities are only known for $N=1,2$. The case $N=1$ is the Ising model where the simplest GKS 2 correlation inequality with $a,b$ of length 1 is due to Griffiths~\cite{Griffiths1}.
The general GKS 2 inequalities for the Ising model were then established by Kelly and Sherman~\cite{KellyS}, hence the now standard GKS acronym.
Further generalizations and improvements were obtained in~\cite{Sherman} and~\cite{Ginibre1}.
The case $N=2$ is the XY or rotator model where the GKS 2 inequalities were proved by Ginibre in~\cite{Ginibre2}. For $N\ge 3$, i.e., when the symmetry group $O(N)$ becomes non-Abelian, proving the GKS 2 inequalities for invariant observables is an open problem. If one considers correlations between non-invariant observables like $\sigma_{i,\ell}$ or particular spin components, there are some results (see~\cite{Lees} and references therein), however they explore a direction which is different from the main focus of this article.
As for the $\mathbb{C}\mathbb{P}^{N-1}$ model, it is trivial for $N=1$, and for any $N\ge 2$ where the symmetry group $U(N)$ (or $SU(N)$) is non-Abelian, the GKS 2 inequalities again represent an open problem.

We now pick up the thread of the previous general discussion of $(X,\mu,\mO) $ systems. In all cases where the GKS 2 inequalities were established, this was done in a coefficient-wise manner in the series expansion of unnormalized correlations with respect to the $J$ couplings.
Since $X$ is compact and the functions involved are continuous and therefore bounded, we
have
\[
\int_X |\mO^a|
\exp\left(\sum_{i=1}^{m}|J_i|\  |\mO(x)^{V_{i\ast}}|\right)
\ {\rm d}\mu(x)<\infty\ .
\]
It is therefore immediate that one can expand the exponential and commute the sum and integral
and write the convergent expansion
\[
\langle\mO^a\rangle_{J,{\rm un}}
=\sum_{\alpha\in\mathbb{N}^m}\frac{J^{\alpha}}{\alpha!}
\langle \mO^{a+\alpha V}\rangle
\ .
\]
In fact, the unnormalized correlations
are entire analytic functions of the $J$ couplings.
In the above equation, we used the standard notation related to multiindices $J^{\alpha}:=J_1^{\alpha_1}\cdots J_{m}^{\alpha_m}$ and $\alpha!:=\alpha_1!\cdots\alpha_m!$. We also used the following elementary identities which provide useful practice with our chosen system of notation:
\[
\prod_{i=1}^{m}\left[\mO(x)^{V_{i\ast}}\right]^{\alpha_i}
=\prod_{i=1}^{m}\left[
\prod_{j=1}^{n}
\mO_j(x)^{V_{i,j}}\right]^{\alpha_i}
=\prod_{j=1}^n\prod_{i=1}^{m}\mO_j(x)^{\alpha_i V_{i,j}}=\mO^{\alpha V}\ .
\]
Indeed, the multiindex $\alpha$ is here considered as a row vector of length $m$ while $V$ is a $m\times n$ matrix and $\alpha V$ simply denotes the matrix product.
After clearing denominators, the GKS 2 inequalities become
\[
\langle\mO^{a+b}\rangle_{J,{\rm un}}\ \langle 1\rangle_{J,{\rm un}}
-\langle\mO^a\rangle_{J,{\rm un}}\ \langle\mO^b\rangle_{J,{\rm un}}
\ge 0\ .
\] 
By imposing that these inequalities hold separately for every coefficient of a (normalized) monomial $\frac{J^{\gamma}}{\gamma!}$, we obtain what we call the coefficient-wise GKS 2 or CGKS 2 inequalities.

\medskip\noindent
{\bf The CGKS 2 inequalities:}
We will say that the system $(X,\mu,\mO)$ satisfies the CGKS 2 collection of inequalities iff
$\forall m\ge 0$, $\forall V\in\mathbb{N}^{m\times n}$,
$\forall a,b\in\mathbb{N}^n$, $\forall\gamma\in\mathbb{N}^{m}$,
we have
\begin{equation}
\sum_{\substack{\alpha,\beta\in\mathbb{N}^m\\\alpha+\beta=\gamma}}
\binom{\gamma}{\alpha}
\left[
\langle\mO^{a+b+\alpha V}\rangle\ \langle\mO^{\beta V}\rangle
-\langle\mO^{a+\alpha V}\rangle\ \langle\mO^{b+\beta V}\rangle
\right]\ge 0\ ,
\label{cgks2ineq}
\end{equation}
where $\binom{\gamma}{\alpha}$ denotes the product of binomials $\prod_{i=1}^{m}
\binom{\gamma_i}{\alpha_i}=\frac{\gamma!}{\alpha!\beta!}$.

Although rather easy to prove for the $O(N)$ and $\mathbb{C}{P}^{N-1}$ models, we can state, for the sake of completeness, the coefficient-wise GKS 1 inequalities as follows.

\medskip\noindent
{\bf The CGKS 1 inequalities:}
We will say that the system $(X,\mu,\mO)$ satisfies the CGKS 1 collection of inequalities iff
$\forall a\in\mathbb{N}^n$,
we have
\[
\langle\mO^a\rangle\ge 0\ .
\]

Clearly, GKS 1 follows from CGKS 1 just as GKS 2 follows from CGKS 2. The advantage of CGKS 2 is that the integrals involved are with respect to the free natural group-invariant measure $\mu$, and this turns the problem into one in representation theory and harmonic analysis on Lie groups and associated homogeneous spaces.
Since this does not seem to have been explicitly formulated as a conjecture, previously in the literature, let us do so and state the following conjecture.

\begin{conjecture}
For all integers $N\ge 3$, the $O(N)$ model satisfies the CGKS 2 inequalities, and for all integers $N\ge 2$, the $\mathbb{C}\mathbb{P}^{N-1}$ model satisfies the CGKS 2 inequalities.
\end{conjecture}

We limited the range of $N$ from below in order to avoid the trivial or settled cases. Although with less confidence than the above conjecture, we also believe that the correct quantification should, in fact, be over any {\it real number} $N\in[1,\infty)$. However, this requires additional definitions related to analytic continuation (see~\cite{BinderR}) which we prefer to leave for a future article.

Using the standard technique of variable duplication, as well as symmetrization over the variables $x,y$, one can rewrite the left-hand side of (\ref{cgks2ineq}) as
\[
\frac{1}{2}\int_{X^2}{\rm d}\mu(x){\rm d}\mu(y)
\ (\mO(x)^a-\mO(y)^a)(\mO(x)^b-\mO(y)^b)\times\prod_{i=1}^{m}\left[
\mO(x)^{V_{i\ast}}+\mO(y)^{V_{i\ast}}
\right]^{\gamma_i}\ .
\] 
The CGKS 2 inequalities are thus a special case of the following inequalities which we call the general Ginibre (GG) inequalities.

\medskip\noindent
{\bf The GG inequalities:}
We will say that the system $(X,\mu,\mO)$ satisfies the GG collection of inequalities iff
$\forall m\ge 0$, $\forall V\in\mathbb{N}^{m\times n}$,
and $\forall (\varepsilon_1,\ldots,\varepsilon_m)\in\{-1,1\}^m$, we have
\[
\int_{X^2}{\rm d}\mu(x){\rm d}\mu(y)
\ \prod_{i=1}^{m}\left[
\mO(x)^{V_{i\ast}}+\varepsilon_i\mO(y)^{V_{i\ast}}
\right]\ge 0\ .
\]

Up to a trivial redefinition of $V$ and $\gamma$, the CGKS 2 inequalities correspond to the situation where only two $\varepsilon$'s are minus signs while all the other are plus signs. We will say that a factor
$\mO(x)^a\pm\mO(y)^a$ has degree $d$ if the length of the multiindex $a$ is equal to $d$. The linear Ginibre inequalities (LG) are the subcollection of the GG
inequalities where all factors have degree one, i.e., $|V_{i\ast}|=1$ for all $i$.
By repeatedly using the elementary identities
\begin{align*}
a_1a_2+b_1b_2 &=\frac{1}{2}(a_1+b_1)(a_2+b_2)+\frac{1}{2}(a_1-b_1)(a_2-b_2)\ ,\\
a_1a_2-b_1b_2 &=\frac{1}{2}(a_1+b_1)(a_2-b_2)+\frac{1}{2}(a_1-b_1)(a_2+b_2)\ ,
\end{align*}
one can break higher degree factors and reduce the proof of the GG inequalities to the particular case of LG inequalities. This is how all the above inequalities were derived by Ginibre~\cite{Ginibre2} for the $O(N)$ model with $N\in\{1,2\}$.
For reasons which will become apparent soon we will introduce even more general inequalities which we call the padded general Ginibre inequalities (PGG). In order to state them we need to define a notion of parity, first in the general $(X,\mu,\mO)$ abstract setting, and then explaining them on the two particular models we have been considering.

Suppose we are given a group homomorphism $\rho:\mathbb{Z}^n\rightarrow (\mathbb{Z}/2\mathbb{Z})^L$, for some integer $L\ge 0$. We think of the image $\rho(e_j)$ of the $j$-th canonical basis vector as the parity check vector for the basic observable $\mO_j$. Therefore $\rho(a)$ for a multiindex $a\in\mathbb{N}^n\subset\mathbb{Z}^n$ is the parity check vector for the monomial $\mO^a$. We will say that $a$ is even iff $\rho(a)=0$. In the case of the $O(N)$ model as described above, we take $L=p$ and for $j$ corresponding to a pair of vertices $(i,i')$, with $1\le i<i'\le p$,
we define $\rho(e_j)$
as the vector with all components equal to 0 except the $i$'-th and $i'$-th components which are set equal to 1.
In the case of the Ising model ($N=1$), we trivially have
\[
\langle\mO^a\rangle=\bbone\{a\ {\rm is\ even}\}\ ,
\]
where $\bbone\{\cdots\}$ is the indicator function of the condition between braces.
For general $N\ge 1$, if one changes one spin $\sigma_i$ to $-\sigma_i$ then the measure
of integration $\mu$ is not affected. The integral $\langle\mO^a\rangle$ is then equal to itself times $(-1)^{\rho_i(a)}$ where $\rho_i(a)\in\mathbb{Z}/2\mathbb{Z}$ is the $i$-th component of the parity check vector $\rho(a)$.
Hence, the expectation $\langle\mO^a\rangle$ vanishes if $a$ is not even.
It is not hard to prove, for any $N\ge 1$, that conversely if $a$ is even then $\langle\mO^a\rangle>0$.
In the case of the $\mathbb{C}\mathbb{P}^{N-1}$ model, the condition of being even is by definition always satisfied. One can take $L=0$ and $\rho$ the trivial unique possible homomorphism, or one can still keep $L=p$, the number of vertices or lattice sites, while defining $\rho$ as the trivial homomorphism. We can now formulate the padded versions of the CGKS 2 and GG inequalities.

\medskip\noindent
{\bf The PCGKS 2 inequalities:}
We will say that the system $(X,\mu,\mO,L,\rho)$ satisfies the PCGKS 2 collection of inequalities iff
$\forall m\ge 0$, $\forall V\in\mathbb{N}^{m\times n}$,
$\forall a,b\in\mathbb{N}^n$, $\forall\gamma\in\mathbb{N}^{m}$, and all even $u\in\mathbb{N}^n$,
we have
\[
\sum_{\substack{\alpha,\beta\in\mathbb{N}^m\\\alpha+\beta=\gamma}}
\binom{\gamma}{\alpha}
\left[\ 
\langle\mO^{u+a+b+\alpha V}\rangle\ \langle\mO^{u+\beta V}\rangle
-\langle\mO^{u+a+\alpha V}\rangle\ \langle\mO^{u+b+\beta V}\rangle
\ \right]\ge 0\ .
\]

\medskip\noindent
{\bf The PGG inequalities:}
We will say that the system $(X,\mu,\mO,L,\rho)$ satisfies the PGG collection of inequalities iff
$\forall m\ge 0$, $\forall V\in\mathbb{N}^{m\times n}$,
$\forall (\varepsilon_1,\ldots,\varepsilon_m)\in\{-1,1\}^m$, and all even $u\in\mathbb{N}^n$
we have
\[
\int_{X^2}{\rm d}\mu(x){\rm d}\mu(y)\ \mO(x)^u\ \mO(y)^u\ 
\prod_{i=1}^{m}\left[
\mO(x)^{V_{i\ast}}+\varepsilon_i\mO(y)^{V_{i\ast}}
\right]\ge 0\ .
\]
If one prefers not to
use the variable duplication trick, then the last inequality can be rewritten as
\begin{equation}
\sum_{\substack{\alpha,\beta\in\mathbb{N}^m\\ \alpha+\beta=\mathbf{1}_m}}
\varepsilon^{\beta}
\langle\mO^{u+\alpha V}\rangle\ \langle\mO^{u+\beta V}\rangle
\ge 0\ ,
\label{PGGnodup}
\end{equation}
where $\mathbf{1}_m:=(1,\ldots,1)\in\mathbb{N}^m$, $\varepsilon^{\beta}=\varepsilon_{1}^{\beta_1}\cdots\varepsilon_{m}^{\beta_m}$, and the binomial is not needed.

Note that restricting to $u$ even is not excessive precaution because, for example,
\[
\frac{1}{16}\sum_{\sigma_1,\sigma_2,\tau_1,\tau_2=\pm 1}
(\sigma_1\sigma_2)(\tau_1\tau_2)[(\sigma_1\sigma_2)-(\tau_1\tau_2)]^2=
-\frac{1}{16}\sum_{\sigma_1,\sigma_2,\tau_1,\tau_2=\pm 1}
[(\sigma_1\sigma_2)-(\tau_1\tau_2)]^2=-2<0\ .
\]
This is a simple Ising example with two lattice sites, namely with $N=1$, $p=2$. The exponent vector $u$ reduces to a single integer equal to one, and $u$ or equivalently the corresponding obervable $\mO=\sigma_1\sigma_2$ is not even.
As noted earlier the CGKS 2 inequalities are particular cases of the GG inequalities, and similarly, it is easy to see that the PGG inequalities include the PCGKS 2 ones as special cases.
Now is a good time to recall that the GG inequalities for the $O(N)$ model with $N\ge 3$ are not true in the level of generality stated above. Indeed, Sylvester~\cite{Sylvester} found examples where the inequalities fail for $N$ large enough. However, his experimental findings do not suggest that the GG inequalities are always false, but rather that they tend to be true if there are not too many $\varepsilon$'s which are minus signs. In particular, he found no counter-example to the CGKS 2 inequalities. 
Given that some GG inequalities (which are not true!) would have to be discarded, an inductive approach to proving CGKS 2, using the $a_1a_2\pm b_1 b_2$ reduction identities or some other idea, would likely benefit from a fresh supply of new inequalities which could be used as intermediates. This is one of our motivations for introducing the padded generalization of CGKS 2 and GG.
Note that the PCGKS 2 inequalities are reminiscent of inequalities considered by Sherman~\cite{Sherman} and Ginibre~\cite[Thm. 2]{Ginibre1} where $u$ is summed over instead of being fixed as in this article. The addition of the padding $u$ is also similar in flavor to the strong subadditive property (as opposed to mere subadditivity) of entropy~\cite{LiebR}.

Another possibility to try to prove CGKS 2 is a bijective approach as with the so-called switching lemma (see, e.g.,~\cite{DuminilC}). Let $K$ be a finite set together with a decomposition into disjoint subsets $K_1,\ldots,K_m$ such that, for each $i$, the cardinality $|K_i|$ of $K_i$ is equal to $\gamma_i$. Then (\ref{cgks2ineq})
can be rewritten as
\begin{equation}
\sum_{A\subset K}
\langle\mO^{a+b+\{A\} V}\rangle\ \langle\mO^{\{K\backslash A\} V}\rangle
\ge
\sum_{A\subset K}
\langle\mO^{a+\{A\} V}\rangle\ \langle\mO^{b+\{K\backslash A\} V}\rangle
\label{bijineq}
\end{equation}
where $\{A\}$ denotes the multiindex of intersection cardinalities $(|A\cap K_1|,\ldots,|A\cap K_m|)$. The inequality would follow if one could produce a bijection $\Psi$ of the power set of $K$ such that for all subsets $A$,
\[
\langle\mO^{a+b+\{A\} V}\rangle \langle\mO^{\{K\backslash A\} V}\rangle
\ge
\langle\mO^{a+\{\Psi(A)\} V}\rangle \langle\mO^{b+\{K\backslash \Psi(A)\} V}\rangle
\ .
\]
For the $N=1$ Ising case, this is easy to do because the correlations are indicator functions of the even condition. Either the right-hand side of (\ref{bijineq})
is zero and there is nothing to prove, or there exists a $C\subset K$ such that $a+\{C\} V$ and $b+\{K\backslash C\} V$ are even.
A bijection which works then is $\Psi(A):=K\backslash(A\Delta C)$ where $\Delta$ denotes the symmetric difference of sets.
Indeed, a quick computation gives
\[
a+\{\Psi(A)\} V=a+b+\{A\}V-(b+\{K\backslash C\}V)-2\{A\}V-2\{A\cap C\}V
\]
and the last three terms obviously are in the kernel of the parity check map $\rho$.
Therefore $\langle\mO^{a+\{\Psi(A)\} V}\rangle=
\langle\mO^{a+b+\{A\} V}\rangle$. One similarly checks $\langle\mO^{b+\{K\backslash\Psi(A)\} V}\rangle=
\langle\mO^{\{A\} V}\rangle$. In order to extend such a switching lemma to $N>1$ one would need a deep understanding of inequalities of the form
\begin{equation}
\langle\mO^a\rangle\ \langle\mO^b\rangle\ge
\langle\mO^c\rangle\ \langle\mO^d\rangle
\label{compareineq}
\end{equation}
where the exponent multiindices $a,b,c,d\in\mathbb{N}^n$ are subject to the constraint $a+b=c+d$. This is a problem analogous to that considered in~\cite{Okounkov}, see also~\cite{DobrovolskaP} for a specific conjecture about products of Schur functions with a flavor similar to (\ref{compareineq}). The difficulty here is to correctly guess a large set $\mathscr{C}$ of quadruples $(a,b,c,d)$ for which
(\ref{compareineq}) is true.
An optimistic yet natural assumption to make on the set $\mathscr{C}$ is that it should be a cone, i.e., invariant under dilation by the same integer $\lambda\in\mathbb{N}_{>0}$. 
It is then natural to take the $\lambda\rightarrow\infty$ limit and see what becomes of the inequality (\ref{compareineq}). The starting point of this article is the author's realization
that for the $O(N)$ and $\mathbb{C}\mathbb{P}^{N-1}$ models, 
modulo some harmless prefactors, the correlations $\langle\mO^{\lambda a}\rangle$ become $K(a)^{-\eta}$ where $K(a)$ is the Kirchhoff polynomial and $\eta$ is a suitable exponent. The latter is $\frac{N-1}{2}$ in the $O(N)$ model case and $N-1$ for the $\mathbb{C}\mathbb{P}^{N-1}$ model.
The simplest PCGKS 2 inequality, for $m=0$, is the statement
\begin{equation}
\langle\mO^{u+a+b}\rangle\ \langle\mO^u\rangle\ge
\langle\mO^{u+a}\rangle\ \langle\mO^{u+b}\rangle
\label{Oinframod}
\end{equation}
or, equivalently, that (\ref{compareineq}) should hold if $b\le c\le a$, where these are componentwise inequalities.
In this article we show that this log-ultramodularity statement is true in the large power limit $\lambda\rightarrow\infty$. Another motivation for the padded versions of the CGKS 2 and GG inequalities, is that padding is necessary for making sense of the large $\lambda$ limit.

\section{Main results}
We will now give precise statements for the main results of this article.
We will start with the large power asymptotics of the $O(N)$ model.
Let $G=(V,E)$ be a simple (undirected) graph with vertex set $V=[p]:=\{1,2,\ldots,p\}$, with $p\ge 2$,
and edge set $E$ identified with a subset of $\{(i,j)\in [p]^2\ |\ i<j\}$. We suppose we are given a collection of integer weights $m=(m_e)_{e\in E}\in\mathbb{N}_{>0}^{E}$.
We will assume that the graph $G$ is connected.
Recall that the Kirchhoff polynomial evaluated on the weights $m$ is given by
\[
K(m):=\sum_{T}\prod_{e\in T} m_e\ ,
\]
where the sum is over all spanning trees $T\subset E$ which connect all $p$ vertices.
With a slight abuse of notation, we will also write $m$ for the matrix $(m_{ij})_{1\le i,j\le p}$ obtained by extending the previous collection as follows. When $i<j$ and $(i,j)\notin E$, we let $m_{ij}=0$. We set all diagonal entries to zero, and finally we force $m$ to be a symmetric matrix, i.e., we let $m_{ij}:=m_{ji}$ if $i>j$.
We then define the vertex degrees $d_i:=\sum_{j=1}^p m_{ij}$ for all $i\in[p]$.
We will assume that $m$ is even, namely that for all $i\in[p]$, we have that $d_i$ is an even integer. This is the same as the parity condition mentioned above but with a different notation and labeling system which is more convenient for stating the following theorem.
We define the integral
\[
\langle \mO^m\rangle:=\int_{(\mathbb{S}^{N-1})^p}{\rm d}\mu(\sigma)
\prod_{1\le i<j\le p}(\sigma_i\cdot\sigma_j)^{m_{ij}}
\] 
over the product of spheres $X=(\mathbb{S}^{N-1})^p$, with respect to $\mu$, the product of $O(N)$-invariant probability measures.

\begin{theorem} (Large power asymptotics for the $O(N)$ model)
We assume $N\in\mathbb{N}_{>0}$, the graph $G$ is connected and the weight collection $m$ is even.
If one multiplies all the weights by $\lambda\in\mathbb{N}_{>0}$ and takes the $\lambda\rightarrow\infty$ limit, 
then we have the asymptotic equivalence
\[
\langle \mO^{\lambda m}\rangle\sim
\left[\frac{2^{\frac{N-1}{2}}\Gamma\left(\frac{N}{2}\right)}{\sqrt{\pi}}\right]^{p-1}\ \lambda^{-\frac{(p-1)(N-1)}{2}}\ K(m)^{-\left(\frac{N-1}{2}\right)}\ .
\]
\label{ONthm}
\end{theorem}

We then have a similar result for the $\mathbb{C}\mathbb{P}^{N-1}$ model. The combinatorial setup is the same except we do not require $m$ to be even, and the integral to be considered is
\[
\langle \mO^m\rangle:=\int_{(\mathbb{S}^{2N-1})^p}{\rm d}\mu(z)
\prod_{1\le i<j\le p}|\langle z_i,z_j\rangle|^{2m_{ij}}
\] 
over the product of complex unit spheres $\mathbb{S}^{2N-1}\subset\mathbb{C}^N$,
with respect to $\mu$, the product of $U(N)$-invariant probability measures.

\begin{theorem} (Large power asymptotics for the $\mathbb{C}\mathbb{P}^{N-1}$ model)
We assume $N\in\mathbb{N}_{>0}$, and the graph $G$ is connected.
If one multiplies all the weights by $\lambda\in\mathbb{N}_{>0}$ and takes the $\lambda\rightarrow\infty$ limit, 
then we have the asymptotic equivalence
\[
\langle \mO^{\lambda m}\rangle\sim
\Gamma(N)^{p-1}\ \lambda^{-(p-1)(N-1)}\ K(m)^{-(N-1)}\ .
\]
\label{CPthm}
\end{theorem}

As a result, one can study the previous inequalities, in the large power limit, using suitable negative powers $\eta$ of the Kirchhoff polynomial. For instance (\ref{Oinframod}) amounts to
\begin{equation}
K(u+a+b)^{-\eta}\ K(u)^{-\eta}\ge K(u+a)^{-\eta}\ K(u+b)^{-\eta}\ .
\label{Kinframod}
\end{equation}
Note that
$K$ is an example of real stable polynomial (see, e.g.,~\cite{Wagner} for a pedagogical review and~\cite{BrandenH} for recent developments), namely it has real coefficients and does not vanish when all the variables are complex numbers with strictly positive imaginary part. It therefore satisfies the Rayleigh property, namely, that $\Delta_{e,f}(K)(x)\ge 0$
for all $x\in\mathbb{R}^{E}$ and all $e,f\in E$, where 
\[
\Delta_{e,f}(K)(x):=\frac{\partial K}{\partial x_e}(x)\ \frac{\partial K}{\partial x_f}(x)
-K(x)\ \frac{\partial^2 K}{\partial x_e\partial x_f}(x)\ .
\]
Working only on $(0,\infty)^E$, where $K$ takes strictly positive values, we have, by elementary calculus,
\[
\log K(u+a+b)+\log K(u)-\log K(u+a)-\log K(u+b)
=\int\limits_{0}^1\ {\rm d}s
\int\limits_{0}^1\ {\rm d}t\ 
\frac{\partial^2}{\partial s\partial t}\log K(u+sa+tb)\le 0
\]
because
\[
\frac{\partial^2}{\partial s\partial t}\log K(u+sa+tb)=
-K(u+sa+tb)^{-2}\ \sum_{e,f\in E}a_e b_f\Delta_{e,f}(K)(u+sa+tb)\ ,
\]
and thanks to the Rayleigh property. In other words (\ref{Kinframod}) is true for any $\eta\in (0,\infty)$.
This, together with the relevant asymptotics theorem, gives us a corollary which says that (\ref{Oinframod}) is true for the $O(N)$ model, at zero coupling, and in the large power limit.
One can prove a plethora of similar corollaries, thanks to the next general result which is the third main result of this article. The idea is that one can use $P(a)^{-\eta}$, with $P$ a suitable polynomial, as a substitute/toy model for the free correlations $\langle\mO^a\rangle$ of spin models, when investigating all the previous collections of inequalities, including the most general PGG ones.
One can also account for parity constraints by using instead the combination $\bbone\{a\ {\rm even}\}P(a)^{-\eta}$ which has the effect of dropping many terms from inequalities like CGKS 2, etc.

Let $q\in\mathbb{N}_{>0}$, and let $A_1,\ldots,A_n$ be $n$ real symmetric positive semidefinite matrices of $q\times q$ format. Suppose that $A_1+\cdots+A_n$ is positive definite and define the polynomial
\[
P(x)={\rm det}(x_1A_1+\cdots x_nA_n)
\]
which is then strictly positive for $x\in(0,\infty)^n$.
A polynomial $P(x)$ that can be represented as above must be real stable~\cite[Lemma 4.1]{Branden1}. Such determinantal polynomials feature prominently in the area of real algebraic geometry which studies hyperbolic polynomials, (see, e.g.,~\cite{Branden2,Kummer,KozhasovMS} and references therein).
We also assume we have at our disposal a parity check homomorphism $\rho:\mathbb{Z}^n\rightarrow (\mathbb{Z}/2\mathbb{Z})^L$ and an associated definition of being even for multiindices $a\in\mathbb{N}^n$, as in the previous section.

\begin{theorem} (The PGG inequalities for stable determinantal polynomials)
For any $r\in\mathbb{N}_{>0}$, the substitute $a\mapsto\bbone\{a\ {\rm even}\}P(a)^{-\frac{r}{2}}$
for the map $a\mapsto\langle\mO^a\rangle$ satisfies all the PGG inequalities.
More precisely, 
$\forall m\ge 0$, $\forall V\in\mathbb{N}^{m\times n}$ such that $\mathbf{1}_m V$ is even,
$\forall (\varepsilon_1,\ldots,\varepsilon_m)\in\{-1,1\}^m$,
and for all even $u\in\mathbb{N}_{>0}^n$,
we have
\[
\sum_{\substack{\alpha,\beta\in\mathbb{N}^m\\ \alpha+\beta=\mathbf{1}_m}}
\bbone\{\alpha V\ {\rm even}\}
\ \varepsilon^{\beta}
\ P(u+\alpha V)^{-\frac{r}{2}} 
\ P(u+\beta V)^{-\frac{r}{2}}
\ge 0\ .
\]
\label{BBthm}
\end{theorem}

By the Matrix-Tree Theorem (see~\cite{Abdesselam} and references therein), the Kirchhoff polynomial $K$ admits a determinantal representation as above and therefore satisfies the hypotheses of Thm. \ref{BBthm}.

\section{Proofs}
\label{proofsec}

Theorems \ref{ONthm} and \ref{CPthm} are rather straightforward exercises in the use of the Laplace method (see, e.g.,~\cite[Ch. 8]{BleisteinH}). For the $O(N)$ model, one uses Fubini to integrate first over $\sigma_2,\ldots,\sigma_p$ and then over $\sigma_1$. However, by $O(N)$-invariance, the integral over $\sigma_2,\ldots,\sigma_p$ is independent of $\sigma_1$ which therefore can be set equal to $(1,0,\ldots,0)^{\rm T}\in\mathbb{R}^N$. Then one can reduce to a fundamental domain of the group $\{1,-1\}^{p-1}$ which flips the direction of individual spins. For example we can reduce to the domain where the first component of each spin $\sigma_2,\ldots,\sigma_p$ is positive. 
The asymptotics are dictated by an isolated maximum where all the spins are equal to $(1,0,\ldots,0)^{\rm T}$ while the Cauchy-Schwarz inequality (and characterization of the equality case) shows that the contribution of the complement of a patch around the maximum is exponentially suppressed. The contribution of the maximum is done as usual by expanding up to second order the logarithm of the integrand. 
One can also integrate over flat space instead of spheres by radially extending integrals over $\mathbb{S}^{N-1}$ to integrals over $\mathbb{R}^{N}$ because the integrand is multihomogeneous of multidegree $(d_2,\ldots,d_p)$.
The determinant of the Hessian, via the Matrix-Tree Theorem, produces the Kirchhoff polynomial. The $\mathbb{C}\mathbb{P}^{N-1}$
model is similar. The fundamental domain after quotienting by the $U(1)^{p-1}$ symmetry now is the domain where the first component of each complex unit vector $z_2,\ldots,z_p$ is in $(0,\infty)$. One again has a unique isolated maximum and the standard Laplace method computation gives the desired result. The upcoming reference~\cite{AbdesselamUW2} will provide detailed proofs of more general asymptotics.

Without further ado, we move on to the proof of Thm. \ref{BBthm}.
For $y\in(0,\infty)^n$, and as in~\cite[Eq. 4.13]{ScottS}, we note that we have a Gaussian convergent integral representation
\[
P(y)^{-\frac{r}{2}}=\int_{\mathbb{R}^{q\times r}}
\exp\left(-\sum_{k=1}^{r}X_{\ast k}^{\rm T}A X_{\ast k}
\right)
\prod_{k=1}^{r}
\prod_{p=1}^{q}
\frac{{\rm d}X_{pk}}{\sqrt{\pi}}
\]
where $A=\sum_{j=1}^{n}y_j A_j$ is symmetric and positive definite.
The integration is over a matrix $X$ of format $q\times r$. The notation $X_{\ast k}$ is for the $k$-th column of $X$.
For the following computations, we will use the abbreviation
\[
{\rm d}X:=\prod_{k=1}^{r}
\prod_{p=1}^{q}
\frac{{\rm d}X_{pk}}{\sqrt{\pi}}
\]
for the integration measure.
We note that
\begin{align*}
\sum_{k=1}^{r}X_{\ast k}^{\rm T}A X_{\ast k} &= 
\sum_{k=1}^{r}
\sum_{p_1,p_2}^{q}
X_{p_1 k}A_{p_1 p_2} X_{p_2 k} \\
 &= {\rm tr}(X^{\rm T}AX) \\
  &= \sum_{j=1}^{n}y_j\ {\rm tr}(X^{\rm T} A_j X)\ .
\end{align*}
Hence,
\begin{align*}
P(y)^{-\frac{r}{2}} &= \int_{\mathbb{R}^{q\times r}}
\exp\left(- \sum_{j=1}^{n}y_j\ {\rm tr}(X^{\rm T} A_j X)
\right)\ {\rm d}X \\
 &= \int_{[0,\infty)^n}e^{-ys^{\rm T}}\ {\rm d}\nu(s)
\end{align*}
where $\nu$ is the push-forward measure on the cone $C:=[0,\infty)^n$
of the measure ${\rm d}X$ via the map
\[
\begin{array}{llll}
\tau: & \mathbb{R}^{q\times r} & \longrightarrow & C \\
 & X & \longmapsto & \tau(X)=\left({\rm tr}(X^{\rm T}A_j X)\right)_{1\le j\le n}\ .
\end{array}
\]

\begin{lemma}\label{keylemma}
For all $v\in(0,\infty)^n$ and for all multiindex $a\in\mathbb{N}^n$, we have
\[
\int_{C^2}{\rm d}\nu(s){\rm d}\nu(t)
\ e^{-v(s+t)^{\rm T}} (t-s)^a\ge 0\ .
\]
\end{lemma}

\smallskip
\noindent{\bf Proof of the lemma:}
By undoing the push--forward by the map $\tau$ we have
\begin{align*}
\mathscr{I} &:= \int_{C^2}{\rm d}\nu(s){\rm d}\nu(t)
\ e^{-v(s+t)^{\rm T}} (t-s)^a\\
 &= \int_{(\mathbb{R}^{q\times r})^2}{\rm d}X\ {\rm d}Y\ 
  \exp\left(
  -{\rm tr}(X^{\rm T}A X) -{\rm tr}(Y^{\rm T}A Y)
  \right)\times
  \prod_{j=1}^{n}\left(
   {\rm tr}(Y^{\rm T}A_j Y) -{\rm tr}(X^{\rm T}A_j X)
  \right)^{a_j}\ ,
\end{align*}
where $A=\sum_{j=1}^{n}v_j A_j$.
Since $A_j$ is symmetric positive semidefinite one can diagonalize it as $A_j=R_j^{\rm T}D_j R_j$ where $R_j$ is orthogonal and $D_j$ is diagonal with nonnegative entries. Therefore, $A_j=B_{j}^{\rm T}B_j$ with $B_j:=\sqrt{D_j} R_j$ and one can write
\[
 {\rm tr}(X^{\rm T}A_j X) =  {\rm tr}((B_j X)^{\rm T} B_j X) 
  = \sum_{p=1}^{q}\sum_{k=1}^{r}[B_j X]_{p,k}^{2}\ ,
\]
where $[B_j X]_{p,k}$ means the $p$-th row and $k$-th column entry of the $q\times r$ matrix $B_j X$, and the $2$ superscript means the square of that entry.
As a result, we get
\begin{align*}
\mathscr{I} &=\int_{(\mathbb{R}^{q\times r})^2}{\rm d}X\ {\rm d}Y\ 
  \exp\left(
  -{\rm tr}(X^{\rm T}A X) -{\rm tr}(Y^{\rm T}A Y)\right) \\
  & \ \ \ \times\prod_{j=1}^{n}
  \left[
  \sum_{p=1}^{q}\sum_{k=1}^{r}
 \left\{ [B_j Y]_{p,k}^{2}-[B_j X]_{p,k}^{2}\right\}
\right]^{a_j}\ .
\end{align*}
By expanding the product, but without breaking the difference terms $\{\cdots\}$, we see that $\mathscr{I}$ decomposes as a sum of integrals of the form
\[
\mathscr{I}_{j,p,k} =\int_{(\mathbb{R}^{q\times r})^2}{\rm d}X\ {\rm d}Y\ 
  \exp\left(
  -{\rm tr}(X^{\rm T}A X) -{\rm tr}(Y^{\rm T}A Y)\right)
 \ \times\prod_{\alpha=1}^{|a|}
 \left\{ [B_{j_\alpha} Y]_{p_\alpha,k_\alpha}^{2}-[B_{j_\alpha} X]_{p_\alpha,k_\alpha}^{2}\right\}\ ,
\]
where $j,p,k$ are now upgraded to multiindices in $[n]^{|a|},[q]^{|a|},[r]^{|a|}$ respectively.
Note the factorization
\[
 [B_{j_\alpha} Y]_{p_\alpha,k_\alpha}^{2}-[B_{j_\alpha} X]_{p_\alpha,k_\alpha}^{2}=
 [B_{j_\alpha}(Y-X)]_{p_\alpha,k_\alpha}\times
  [B_{j_\alpha}(Y+X)]_{p_\alpha,k_\alpha}
\]
which prompts the change of variables $(Y,X)\rightarrow(\widetilde{Y},\widetilde{X})$ given by
\[
\left\{
\begin{array}{ccc}
\widetilde{Y} & = & \frac{1}{\sqrt{2}}(Y-X)\ , \\
\widetilde{X} & = & \frac{1}{\sqrt{2}}(Y+X)\ ,
\end{array}
\right.
\]
and which has a Jacobian equal to $1$.
A quick computation shows 
${\rm tr}(X^{\rm T}A X) +{\rm tr}(Y^{\rm T}A Y)=
{\rm tr}(\widetilde{X}^{\rm T}A \widetilde{X}) +
{\rm tr}(\widetilde{Y}^{\rm T}A \widetilde{Y})$ which immediately implies
\[
\mathscr{I}_{j,p,k} =2^{|a|}\left[
\int_{\mathbb{R}^{q\times r}}{\rm d}\widetilde{X}
\exp\left(-{\rm tr}(\widetilde{X}^{\rm T}A \widetilde{X})\right)
\prod_{\alpha=1}^{|a|}
[B_{j_\alpha} \widetilde{X}]_{p_\alpha,k_\alpha}
\right]^2
\ge 0\ ,
\]
and the lemma follows. \qed

We now pick up the thread of the proof of Thm. \ref{BBthm} and introduce the notation
\[
\Theta:=
\sum_{\substack{\alpha,\beta\in\mathbb{N}^m\\ \alpha+\beta=\mathbf{1}_m}}
\bbone\{\alpha V\ {\rm even}\}\ 
\varepsilon^{\beta}\ 
P(u+\alpha V)^{-\frac{r}{2}}\  
P(u+\beta V)^{-\frac{r}{2}}\ ,
\]
for the quantity we must show is nonnegative.
Under the hypothesis that $\mathbf{1}_m V$ is even, we have that $\alpha V$ is even iff so is $\beta V$. We can thus insert a $\bbone\{\beta V\ {\rm even}\}$ factor
in the formula for $\Theta$.
We will take care of such indicator functions by writing
\[
\bbone\{a\ {\rm even}\} =\frac{1}{2^L}\sum_{g_1,\ldots,g_L=\pm 1}
g_{1}^{\rho_1(a)}\cdots g_{L}^{\rho_L(a)} 
 = \int_G {\rm d}\omega(g)\ g^{\rho(a)}\ ,
\]
where $G$ is the group $(\{1,-1\}^L,\times)$, $\omega$ is the normalized Haar measure on $G$, and $g^{\rho(a)}$ denotes $g_{1}^{\rho_1(a)}\cdots g_{L}^{\rho_L(a)}$. When writing $(\pm 1)^k$, it does not matter of course if we think of $k$ as an element of $\mathbb{Z}$ or of $\mathbb{Z}/2\mathbb{Z}$.
We can now rewrite the quantity of interest as
\[
\Theta=
\sum_{\substack{\alpha,\beta\in\mathbb{N}^m\\ \alpha+\beta=\mathbf{1}_m}}
\varepsilon^{\beta}
\int_{G^2}{\rm d}\omega(g)\ {\rm d}\omega(h)
\int_{C^2}{\rm d}\nu(s)\ {\rm d}\nu(t)\ 
g^{\rho(u+\alpha V)}h^{\rho(u+\beta V)}
\ e^{-(u+\alpha V)s^{\rm T}-(u+\beta V)t^{\rm T}}\ ,
\]
\[
=\int_{G^2}{\rm d}\omega(g)\ {\rm d}\omega(h)
\int_{C^2}{\rm d}\nu(s)\ {\rm d}\nu(t)
\ e^{-us^{\rm T}-ut^{\rm T}}
\times
\left(
\sum_{\substack{\alpha,\beta\in\mathbb{N}^m\\ \alpha+\beta=\mathbf{1}_m}}
\varepsilon^{\beta}
g^{\rho(\alpha V)}h^{\rho(\beta V)}
e^{-\alpha V s^{\rm T}-\beta V t^{\rm T}}
\right)\ ,
\]
since $\rho$ is a homomorphism and $\rho(u)=0$ by hypothesis.
Note that $\rho(\alpha V)=\rho\left(\sum_{i=1}^{m}\alpha_i V_{i\ast}\right)=
\sum_{i=1}^{m}\alpha_i\rho(V_{i\ast})$, where it again does not matter if we think $\alpha_i\in\mathbb{Z}$ (setting of $\mathbb{Z}$-modules) or $\alpha_i\in\mathbb{Z}/2\mathbb{Z}$ (setting of $\mathbb{Z}/2\mathbb{Z}$-vector spaces).
We then have the easy rewriting
\[
g^{\rho(\alpha V)} 
 =  \prod_{i=1}^{m} \left(
g^{\rho(V_{i\ast})} 
 \right)^{\alpha_i}\ .
\]
We also have
\[
e^{-\alpha V s^{\rm T}}=e^{-\sum_{i=1}^{m}\alpha_i V_{i\ast}s^{\rm T}}
= \prod_{i=1}^{m} \left(
e^{-V_{i\ast}s^{\rm T}} 
 \right)^{\alpha_i}\ .
\]
From these elementary identities, we deduce the factorization
\[
\sum_{\substack{\alpha,\beta\in\mathbb{N}^m\\ \alpha+\beta=\mathbf{1}_m}}
\varepsilon^{\beta}
g^{\rho(\alpha V)}h^{\rho(\beta V)}
e^{-\alpha V s^{\rm T}-\beta V t^{\rm T}}
=\prod_{i=1}^{m}
\left[
g^{\rho(V_{i\ast})}e^{-V_{i\ast} s^{\rm T}}+\varepsilon_i
h^{\rho(V_{i\ast})}e^{-V_{i\ast} t^{\rm T}}
\right]\  .
\]
We now make use of the identity for hyperbolic functions
\[
\varepsilon_1 e^{x_1}+\varepsilon_2 e^{x_2}=e^{\frac{x_1+x_2}{2}}\left[
(\varepsilon_1+\varepsilon_2)\cosh\left(\frac{x_1-x_2}{2}\right)
+(\varepsilon_1-\varepsilon_2)\sinh\left(\frac{x_1-x_2}{2}\right)
\right]\ ,
\]
which holds for any signs $\varepsilon_1,\varepsilon_2\in\{1,-1\}$
and $x_1,x_2\in\mathbb{R}$.
We then have
\begin{align*}
\Theta= &  \int_{G^2}{\rm d}\omega(g)\ {\rm d}\omega(h)
\int_{C^2}{\rm d}\nu(s)\ {\rm d}\nu(t)
\ \ e^{-(u+\frac{1}{2}\mathbf{1}_m V)(s+t)^{\rm T}} \\
 & \times\prod_{i=1}^{m}
 \left[
 \left(g^{\rho(V_{i\ast})}+\varepsilon_i h^{\rho(V_{i\ast})}\right)
 \cosh\left(\frac{V_{i\ast}(t-s)^{\rm T}}{2}\right)
 +\left(g^{\rho(V_{i\ast})}-\varepsilon_i h^{\rho(V_{i\ast})}\right)
 \sinh\left(\frac{V_{i\ast}(t-s)^{\rm T}}{2}\right)
 \right]\ .
\end{align*}
We expand the product and separate the integrals over $G^2$ and $C^2$ which gives
\[
\Theta=\sum_{I\subset[m]}\Theta_{G,I}\Theta_{C,I}\ ,
\]
where
\[
\Theta_{G,I}:=\int_{G^2}{\rm d}\omega(g)\ {\rm d}\omega(h)\ 
\prod_{i\in I}
 \left(g^{\rho(V_{i\ast})}+\varepsilon_i h^{\rho(V_{i\ast})}\right)
 \times
 \prod_{i\in[m]\backslash I}
 \left(g^{\rho(V_{i\ast})}-\varepsilon_i h^{\rho(V_{i\ast})}\right)\ ,
\]
and
\[
\Theta_{C,I}:=\int_{C^2}{\rm d}\nu(s)\ {\rm d}\nu(t)
\ e^{-(u+\frac{1}{2}\mathbf{1}_m V)(s+t)^{\rm T}}
\prod_{i\in I}\cosh\left(\frac{V_{i\ast}(t-s)^{\rm T}}{2}\right)
\times
\prod_{i\in[m]\backslash I}\sinh\left(\frac{V_{i\ast}(t-s)^{\rm T}}{2}\right)\ .
\]
Notice that $\Theta_{G,I}\ge 0$ since this is the Ginibre inequality~\cite{Ginibre2} for the Ising model on the set of sites $\{1,\ldots,L\}$. So what remains is to show that the $\Theta_{C,I}$ are nonnegative.
We next insert the series expansions
\[
\cosh x=\sum_{\gamma=0}^{\infty}\frac{x^{2\gamma}}{(2\gamma)!}\ \ ,\ \ {\rm and}\ \ 
\sinh x=\sum_{\gamma=0}^{\infty}\frac{x^{2\gamma+1}}{(2\gamma+1)!}\ ,
\]
which (luckily!) come with nonnegative coefficients, and commute sums and integrals.
Hence,
\[
\Theta_{C,I}=\sum_{\gamma\in\mathbb{N}^n}
\prod_{i\in I}\frac{1}{(2\gamma_i)!}
\times
\prod_{i\in [m]\backslash I}\frac{1}{(2\gamma_i+1)!}
\times\Theta_{C,I,\gamma}\ ,
\]
where
\[
\Theta_{C,I,\gamma}:=
\int_{C^2}{\rm d}\nu(s)\ {\rm d}\nu(t)
\ e^{-(u+\frac{1}{2}\mathbf{1}_m V)(s+t)^{\rm T}}
\prod_{i\in I}\left(\frac{V_{i\ast}(t-s)^{\rm T}}{2}\right)^{2\gamma_i}
\times
\prod_{i\in[m]\backslash I}\left(\frac{V_{i\ast}(t-s)^{\rm T}}{2}\right)^{2\gamma_i+1}\ .
\]
In order to justify the (Fubini) exchange of sum and integrals, one can note the following. We trivially have $\cosh|x|\le e^{|x|}$ and $\sinh |x|\le e^{|x|}$ for all $x\in\mathbb{R}$, as well as the bound
\begin{equation}
|V_{i\ast}(t-s)^{\rm T}|\le V_{i\ast}t^{\rm T}+  V_{i\ast}s^{\rm T}
\label{Fubinijustif}
\end{equation}
because $V,t,s$ have nonnegative entries.
We can replace all expressions $V_{i\ast}(t-s)^{\rm T}$ by the corresponding absolute values $|V_{i\ast}(t-s)^{\rm T}|$,
in the formula for $\Theta_{C,I}$, and do the expansion and (Tonelli) exchange of sum and integral. The result is convergent because after using the bound (\ref{Fubinijustif})
and thus consuming the good factor $e^{-(\frac{1}{2}\mathbf{1}_m V)(s+t)^{\rm T}}$, we still have $e^{-u(s+t)^{\rm T}}$, with $u\in(0,\infty)^n$, which ensures integrability.
Finally, since $V$ has nonnegative entries, we can further expand the products in the formula for $\Theta_{C,I,\gamma}$, without breaking differences $t-s$, 
and Lemma \ref{keylemma} immediately shows $\Theta_{C,I,\gamma}\ge 0$, and we are done. \qed

\section{Some open problems}
The work done in this article raises a host of questions for which we do not yet know the answers. Here is a selection.

\medskip\noindent{\bf Problem 1:}
For the $O(N)$ or $\mathbb{C}{P}^{N-1}$ models, does the inequality (\ref{Oinframod}) hold for $u\neq 0$?
For the $O(N)$ model, with $u=0$, and any $N\in \mathbb{N}_{>0}$, the inequality was proved by Herbst and Baumgartner~\cite{Herbst,Baumgartner}.

\medskip\noindent{\bf Problem 2:}
For the XY model, or $O(2)$ model, the GG inequalities were proved by Ginibre~\cite{Ginibre2}. What about the padded generalizations given by the PGG inequalities?

\medskip\noindent{\bf Problem 3:}
In the light of investigations of spin models with non-integer number of components $N$, as in~\cite{BinderR}, it would be interesting to see if Thm. \ref{BBthm} still holds for $P^{-\eta}$
where $\eta$ is any positive real number instead of being restricted to half integers.
In our proof, the limitation comes from Lemma \ref{keylemma} which crucially relies on the Gaussian integral representation used for $P^{-\eta}$. Of course, the proof with the change of variables $(Y,X)\rightarrow(\widetilde{Y},\widetilde{X})$ belongs to the theme around the Herschel-Maxwell Theorem which characterizes Gaussian random variables and its strengthening by Bernstein (see~\cite{Bryc} for a thorough presentation). One could also have used 
Ginibre's Hermite polynomial example~\cite[Example 5]{Ginibre2}.
Note that the measure $\nu$ used in the integral representation is the inverse Laplace transform of $P^{-\frac{r}{2}}$. The positivity of $\nu$ helps but is not essential for our needs.
One can start with an arbitrary temperate Schwartz distribution $\nu\in \mathscr{S}'(\mathbb{R}^n)$ with support in $C=[0,\infty)^n$, instead of a positive measure, and yet formulate the same positivity statement as in Lemma \ref{keylemma}.
Is there a characterization of Schwartz distributions $\nu$ for which the statement of Lemma \ref{keylemma} holds?
Note that the Laplace transform
\[
\mathscr{L}(x):=\int_{C}e^{-xs^{\rm T}}\ {\rm d}\nu(s)
\]
(with the integral to be interpreted in the sense of distributions) is well defined and analytic in the tube $(0,\infty)^{n}+i\mathbb{R}^n$ (see, e.g.,~\cite[\S{II.2}]{Simon}).
For $\mathscr{L}=P^{-\eta}$, and in the articles~\cite{ScottS,KozhasovMS}, the main focus is the linear positivity requirement: $\forall a\in\mathbb{N}^n$, $\forall x\in (0,\infty)^{n}$,
\[
(-1)^{|a|}\ \partial_x^a\mathscr{L}(x)\ge 0\ .
\]
This requirement severely restricts the allowed exponents $\eta$. Our Lemma \ref{keylemma}
rather is a bilinear positivity requirement: $\forall a\in\mathbb{N}^n$, $\forall x\in (0,\infty)^{n}$,
\[
(\mathscr{L},\mathscr{L})_a(x)\ge 0\ ,
\]
where
\begin{align*}
(\mathscr{L},\mathscr{L})_a(x) &:= \sum_{\substack{b,c\in\mathbb{N}^n\\ b+c=a}}
\binom{a}{b} (-1)^{|c|}\ \partial_x^b\mathscr{L}(x)\ \partial_x^c\mathscr{L}(x) \ , \\
 &= \left. (\partial_x-\partial_y)^a \mathscr{L}(x)\mathscr{L}(y)\right|_{y:=x}\ ,
\end{align*}
which is a Hirota bilinear operator (see~\cite{Hirota}).
For $|a|=2$ (only even lengths are interesting), it is easy to see that the positivity holds, for any $\eta\in(0,\infty)$, because $P$ satisfies the Rayleigh property.
Given the origins of the theory of stable polynomials in classical invariant theory (CIT)~\cite{Grace}, it may help to express the above positivity in an explicitly $SL_{2}^{\times n}$-covariant manner. See~\cite{Leake} for related work aimed at reinterpreting the methods of Borcea and Br\"and\'en in terms of CIT.
In principle, all one has to do is homogenize/dehomogenize, but this is not as easy as it sounds. See, e.g.,~\cite{AbdesselamC} for an example where such a procedure can produce, from a simple nonhomogeneous construction, a rather complicated homogeneous $SL_2$-covariant. A possible way to relate Hirota operators to CIT transvectants was indicated by Olver~\cite[p. 89]{Olver}. These are also close relatives of Rankin-Cohen brackets, here, for Hilbert modular forms~\cite{ChoieKR}.

\medskip\noindent{\bf Problem 4:} Can one understand the bilinear positivity of the previous problem for the true correlations $\langle\mO^a\rangle$ rather than the toy model $P^{-\eta}$?
For (an exceedingly trivial) example, consider the $O(N)$ model with $p=2$, i.e., two lattice sites.
It is easy to compute, with $x\in\mathbb{N}$, the following result, with an unmistakable Mellin-Barnes flavor,
\[
\mathscr{L}(x):=\langle\mO^{2x}\rangle=\frac{\Gamma\left(
\frac{N}{2}\right)\Gamma\left(x+\frac{1}{2}
\right)}{\Gamma\left(\frac{1}{2}\right)\Gamma\left(x+\frac{N}{2}\right)}\ .
\]
One clearly sees the possibility to analytically continue with respect to the number of components $N$ and also the observable exponent $x$.
Using the Bromwich formula, one easily gets the inverse Laplace transform ${\rm d}\nu(s)=f(s){\rm d}s$ where
\[
f(s)=\frac{\Gamma\left(\frac{N}{2}\right)}{\Gamma\left(\frac{1}{2}\right)\Gamma\left(
\frac{N-1}{2}\right)}
\ e^{-\frac{s}{2}}\left(1-e^{-s}\right)^{\frac{N-3}{2}}\ .
\]
In essence, the $\lambda\rightarrow \infty$ large power limit investigated in this article gives information on the behavior of $f$ at the origin. Namely, it gives access to the scaling limit
\[
\lim\limits_{\lambda\rightarrow\infty}\lambda^{\frac{N-3}{2}}f\left(\frac{s}{\lambda}\right)\ .
\]
Considering observable exponents which are non-integers may be more tractable for the $\mathbb{R}\mathbb{P}^{N-1}$ model where the basic observables are nonnegative by construction. Otherwise, one may need discrete Hirota operators as developed for the needs of the theory of quantum integrable systems (see, e.g.,~\cite{KricheverLWZ}).

\bigskip
\noindent{\bf Acknowledgements:}
{\small
The author thanks Ira Herbst for explaining his heat kernel proof~\cite{Herbst} of the GKS 2 inequality at zero coupling. The author thanks Pavlo Pylyavskyy for explaining the notion of alcoved polytopes involved in~\cite{DobrovolskaP}. The author thanks Gennady Uraltsev and Joe Webster for the collaboration~\cite{AbdesselamUW1,AbdesselamUW2} which will show the scope of Thm. \ref{BBthm} extends far beyond the examples provided by the $O(N)$ and $\mathbb{C}\mathbb{P}^{N-1}$ models.
}

\end{document}